\documentclass[aps,prl,amsmath,amssymb,reprint]{revtex4-1}
\usepackage[utf8]{inputenc} 

\usepackage{graphicx} 
\usepackage{color}

\usepackage{booktabs} 
\usepackage{array} 
\usepackage{paralist} 
\usepackage{verbatim} 
\usepackage{subfig} 
\usepackage{hyperref}

\makeatletter
\newcommand*{\rom}[1]{\expandafter\@slowromancap\romannumeral #1@}
\makeatother

\begin{document}
\title{ Born-Oppenheimer Dynamics, Electronic Friction, and the Inclusion of  Electron-Electron Interactions}
\author{Wenjie Dou,$^{1}$ Gaohan Miao,$^{1}$ and Joseph E. Subotnik$^{1, 2}$ }
\affiliation{$^{1}$Department of Chemistry, University of Pennsylvania, Philadelphia, Pennsylvania 19104, USA \\
$^{2}$Stanford PULSE Institute, SLAC National Accelerator Laboratory, Menlo Park, California 94025, USA}

\begin{abstract}

We present a universal expression for the electronic friction as felt by a set of classical nuclear degrees of freedom (DoF's) coupled to a manifold of quantum electronic DoF's; no assumptions are made regarding the nature of the electronic
Hamiltonian and electron-electron repulsions are allowed. Our derivation is based on a quantum-classical Liouville equation (QCLE) for the coupled electronic-nuclear motion, followed by  an adiabatic approximation whereby electronic transitions are assumed to equilibrate faster than nuclear movement. 
The resulting form of friction is completely general, but does reduce to previously published expressions for the
quadratic Hamiltonian (i.e. Hamiltonians without electronic correlation).
At equilibrium, the second fluctuation-dissipation theorem is satisfied and the frictional matrix is symmetric. To demonstrate
the importance of electron-electron correlation, we study electronic friction within the Anderson-Holstein model,
where a proper treatment of electron-electron interactions shows 
signatures of a Kondo resonance and a mean-field treatment is completely inadequate.

\end{abstract} 
 
\maketitle

\textit{Introduction.---}
The Born-Oppenheimer (BO) approximation is probably the most important framework underlying modern physics and chemistry. According to the BO approximation, for a system of nuclei and electrons, we split up the total Hamiltonian into the nuclear kinetic energy ($\hat{T}_{nuc}$) and the electronic Hamiltonian $\hat H$:
\begin{eqnarray}
\label{eq:bo1}
\hat{H}_{tot} &=& \hat T_{nuc} + \hat H \\
\label{eq:bo2}
\hat H &=& \hat{T}_{e} + \hat{V}_{ee} + \hat{V}_{nn}+ \hat{V}_{en} 
\end{eqnarray}
\noindent The electronic Hamiltonian $\hat H$ includes the electronic kinetic energy $\hat{T}_{e}$, 
electron-electron repulsion  ($\hat{V}_{ee}$),
nucleus-nucleus repulsion  ($\hat{V}_{nn}$), and nucleus-electron attraction ($\hat{V}_{en}$).
According to the BO approximation, we diagonalize $\hat H$ and propagate all nuclear motion along a single eigenvalue of $\hat H$, which is
called an adiabatic state; for a large system in the condensed phase, that special state is usually chosen to be the ground state.  

When applying the BO approximation, one must use caution because BO approximation is strictly valid
only when nuclear motion is infinitesimally slow, and there are very well known instances where BO approximation breaks down: 
see, e.g., the recent work of Wodtke on vibrational relaxation and electron transfer at metal surfaces \cite{science2000, bunermann2015electron}. Furthermore, nonadiabatic effects in molecular electronics are known to account for a huge number of interesting phenomena including heating \cite{heating, prbqme}, instability \cite{runaway, thossinstabilities}, and inelastic scattering effects \cite{inelastic, mishaScience2008}.  
Thus, for many experiments, theory must go beyond the BO approximation. 

The goal of the present paper is to show that, each 
and every time one invokes the BO approximation in the condensed phase -- 
provided there is a continuous manifold of electronic states that relax quickly --
there is a single, unique Fokker-Planck equation guiding the nuclear dynamics \cite{smith1993electronic}.
In other words, if one is considering a system with a manifold of electronic states and one wishes to treat the nuclei classically,
BO dynamics should always be propagated with a well-defined frictional damping term and corresponding random force.
The universal expressions for such friction and random force are presented below and should be applicable to  many dynamical scenarios: e.g.,
nuclei scattering off metal surfaces \cite{wodtkereview2004}, atoms vibrating within metal surfaces \cite{ibach2013electron}, molecules relaxing when tethered to photo-excited metals \cite{zhu2002electron},
and molecules stretching and contracting when experiencing a current (and sitting between two metal contacts) \cite{tao2006electron, bruot2012mechanically}.

\textit{Previous Results.---}
The notion that the BO approximation can sometimes lead to friction is not new. In particular, 
such a friction was identified long ago in the context of molecular motion at metal surfaces, 
where a manifold of electronic states is clearly present and  can lead to so-called ``electronic friction.''

One early derivation of electronic friction is due to Head-Gordon and Tully (HGT) \cite{FrictionTully, ryabinkin2016mixed}, who 
followed the dynamics of electrons and nuclei with Ehrenfest dynamics. At zero temperature, and without any electron-electron interactions in the Hamiltonian, they derived the following functional form for the electronic friction:
\begin{eqnarray}\label{eq:mf} 
\gamma_{\alpha \nu}  &=& \pi \hbar\: \sum_{pq}   \langle \phi_p |  \partial_{\alpha} \hat V_{SCF}  | \phi_q \rangle   \langle \phi_q |  \partial_{\nu} \hat V_{SCF}  | \phi_p \rangle  \nonumber \\ 
&& \times \delta(\epsilon_F - \epsilon_q) \delta(\epsilon_F - \epsilon_p) 
\end{eqnarray}
Here $\epsilon_p$ and  $\phi_p$  are, respectively, the energies and orbitals that diagonalize $\hat V_{SCF}$,
the one-electron self-consistent potential. $\epsilon_F$ is the Fermi level, and $\alpha$ (or $\nu$) indices nuclear DoF's. To date, this mean-field form of electronic friction has been applied to many systems using ab
initio electronic structure theory (e.g., DFT) \cite{shenvi:2009:iesh,  tullyappPCCP2011, tullyfrictionexpPRL1996, PhysRevB.94.115432,
Luntz2008143,
luntz2006femtosecond,
luntz2005adiabatic}, and Langevin dynamics on a metal surface with DFT potentials has become a standard tool.

We emphasize, however,  that Eq. \ref{eq:mf} is based on the assumption of independent (either free or mean-field) electrons at equilibrium (i.e. with only one metallic lead).
That being said, this form for electronic friction is consistent with the one-dimensional rates for vibrational relaxation previously published by Persson and Hellsing \cite{PhysRevLett.49.662, hellsing1984electronic}
and others \cite{ PERSSON1980175,  PhysRevB.6.2577}.
Several other research groups have also identified
the same effective friction tensor \cite{paperIV} using different methodologies, some using influence functionals \cite{brandbyge} and some using non-equilibrium Green's functions (NEGF) \cite{lvPRBfriction, Mozyrsky}. The effects of 
non-Condon terms have also been considered \cite{Mizielinski2005, dou2017electronic, mishaPRBfriction} at finite temperature.


More recently, 
von Oppen and co-workers have provided an explicit form for the electronic friction using a non-equilibrium
Green's function and scattering matrix approach \cite{beilstein} plus an explicit adiabatic expansion in terms of nuclear velocity.
The resulting expression is valid both in and out of equilibrium, e.g. for a molecule sitting between two metals with a current running through it.  However, again one must assume either
 a quadratic Hamiltonian without electron-electron repulsion \cite{beilstein} or a mean-field electronic Hamiltonian \cite{vonOppenJCP2013}.
(In a forthcoming article, we will show that the von Oppen and HGT expressions are identical at equilibrium \cite{HGTandvonOppen2017}.)


In the end, many distinct approaches for electronic friction can be found in the literature, most assuming electrons at equilibrium and almost all assuming non-interacting electrons.
%
However, for a realistic description of molecular electronic structure, even if we ignore the effects of an external voltage, 
it is well-known that one cannot waive away electron-electron interactions \cite{martin2016interacting}. 
Despite decades of work on electronic friction, we believe the most general expression for electronic friction (that includes electron-electron correlation) arguably still belongs to Suhl and co-workers \cite{PhysRevB.11.2122}.  
In Ref.  \cite{PhysRevB.11.2122}, the authors
conjectured (without proof) a form for electronic friction that does actually include electron-electron correlation (see below).
A few years ago, Daligault and Mozyrsky \cite{daligault2007ion} 
successfully derived Suhl's conjecture at equilibrium
with a proper random force for the first time, although they did not investigate electron-electron interactions
explicitly. Most importantly, we emphasize that Refs. \cite{PhysRevB.11.2122, daligault2007ion} 
are limited to an electronic system in equilibrium.

\textit{Outline.---}
Given $(i)$ how many important effects break the BO approximation, $(ii)$ how little attention has been paid to the
effects of electron-electron correlation on the friction tensor \cite{yoshimori1982friction, LangrethPRB1998, LangrethPRB1999}, and $(iii)$ how many experiments routinely 
apply voltages to metals with molecules nearby,
{\em  the goal of the present letter is to derive one universal expression for electronic friction (with a random force) that is valid both with and without
electron-electron interactions and in and out of equilibrium.}
 We will also show that this 
universal friction reduces to the HGT model for the case
of a quadratic Hamiltonian (i.e. free electrons) at equilibrium.
At equilibrium, our final expression 
matches Suhl's expression \cite{PhysRevB.11.2122,daligault2007ion} and can be understood easily 
through the lens of linear response and correlation functions. However, we emphasize that, in contrast with Refs. \cite{PhysRevB.11.2122,daligault2007ion},
 our derived Fokker-Planck equation 
is valid out of equilibrium so that the effects of non-equilibrium initial conditions can be analyzed.  We will show that 
the second fluctuation-dissipation theorem is satisfied only at equilibrium.

Lastly, to demonstrate just why electron-electron interactions are so important for nonadiabatic effects,
we will study the electronic friction  tensor for the Anderson-Holstein model in the limit of reasonably large $U$.
Here, we will show that a mean-field treatment of electron-electron interactions (as in Eq. \ref{eq:mf})
 can yield a qualitatively incorrect picture
of electronic friction. Given the current push to extend correlated electronic structure methods (beyond mean-field theory)
to extended systems \cite{mcclain2017gaussian, deslippe2012berkeleygw}, the present letter should be immediately useful for describing coupled nuclear-electron motion
in the condensed phase.


\textit{Theory.---}  
Consider the very general Hamiltonian in Eqs. \ref{eq:bo1}-\ref{eq:bo2} and 
let us make a temperature-dependent BO approximation, whereby we assume that the nuclei are propagated along
a Boltzmann average of the Born-Oppenheimer adiabatic surfaces:
\begin{eqnarray} \label{eq:meanF}
F_{\alpha} &=& -  tr_e \big(\partial_{\alpha} \hat{H} \hat{\rho}_{ss} \big) 
\end{eqnarray}
Here, $\hat{\rho}_{ss}$ is a steady-state electronic density matrix.
At equilibrium and in the limit of zero temperature, $\hat{\rho}_{ss} = \left|g\right>\left<g \right|$ and the force $F_{\alpha}$ is simply the ground-state force, which is the more standard 
BO approximation. $tr_e$ implies tracing over all electronic DoF's. 

As shown in the Supplemental Material (SM, Sec. I),
starting with the quantum-classical Liouville equation (QCLE) \cite{KapralQCLE1999, kapral:QCLEreview} and assuming that 
electronic motion is much faster than nuclear motion, 
we derive a Fokker-Planck equation for the nuclear motion in the same spirit as a  Mori-Zwanzig projection \cite{mori1965transport, PhysRev.124.983, berne2000dynamic,romero1989relaxation, romero1989theory, toutounji2001subsystem}. 
We denote the nuclear phase space density  as
$\mathcal{A}(\bold{R}, \bold{P})$. The explicit Fokker-Planck equation reads:
\begin{eqnarray} \label{eq:FP}
\partial_t \mathcal{A}  &=& - \sum_{\alpha} \frac {P_{\alpha} } {m_{\alpha} }  \partial_{\alpha} \mathcal{A}   - \sum_{\alpha} F_{\alpha}  \frac { \partial  \mathcal{A}   } { \partial P_\alpha }  \nonumber \\
&& +   \sum_{\alpha\nu} \gamma_{\alpha\nu}  \frac { \partial  } { \partial P_\alpha } ( \frac {P_{\nu} } {m_{\nu} }   \mathcal{A} ) + \sum_{\alpha\nu} \bar D^{S}_{\alpha\nu}  \frac { \partial^2  \mathcal{A}} { \partial P_\alpha \partial P_\nu } 
\end{eqnarray}
Thus, all BO trajectories should be Langevin dynamics with an
electronic friction and a random force:
\begin{eqnarray} \label{eq:Langevin}
- m_{\alpha} \ddot R_{\alpha} = - F_{\alpha} + \sum_{\nu} \gamma_{\alpha \nu} \dot  R_{\nu}  - \zeta_{\alpha}
\end{eqnarray} 


The electronic  friction has the following simple form in the time domain (expressed in terms of correlation functions)
\begin{eqnarray} \label{eq:friction2}
\gamma_{\alpha \nu} =- \int_0^\infty dt \:  tr_e \Big( \partial_{\alpha} \hat H e^{- i \hat H t/\hbar} \partial_{\nu} \hat \rho_{ss}  e^{ i \hat H t/\hbar } \Big) 
\end{eqnarray}
or the following form in the energy domain (expressed in terms of Greens functions)
\begin{eqnarray} \label{eq:frictionGA}
\gamma_{\alpha \nu} 
&=& - \hbar \int_{-\infty}^\infty  \frac{d\epsilon}{2\pi}  \:  tr_e \Big( \partial_{\alpha} \hat H \hat G^R(\epsilon) \partial_{\nu} \hat \rho_{ss} \hat G^A (\epsilon) \Big) 
\end{eqnarray}
Here we have defined the many-body (as opposed to one-body) retarded and advanced Green's functions $\hat G^{R/A}(\epsilon) =(\epsilon- \hat H \pm i\eta)^{-1}$.  
These expressions prove that the friction must always be real,
$\gamma_{\alpha \nu} = \gamma_{\alpha \nu}^{\dagger}$.

For the random force, at steady state and in the Markovian limit,  we find:
\begin{eqnarray} \label{eq:corrTimeQCLE}
\bar D^S_{\alpha \nu}   
& = & \frac12  \int_0^\infty dt \:    tr_e \big(   e^{i\hat Ht/\hbar}   \delta \hat F_{\alpha}   e^{-i\hat H t/\hbar}  ( \delta \hat F_{\nu}  \hat \rho_{ss} + \hat \rho_{ss}  \delta \hat F_{\nu} )  \big) \nonumber \\
\delta \hat F_{\alpha} &\equiv& - \partial_{\alpha} \hat H +  tr_e(\partial_{\alpha} \hat H \hat{\rho}_{ss} )  \label{eq:Ranforce}
\end{eqnarray}
so that 
\begin{eqnarray} \label{eq:FFc}
\frac{1}{2} \left(\left<\zeta_{\alpha}(t) \zeta_{\nu}(t') \right>   + \left<\zeta_{\nu}(t) \zeta_{\alpha}(t')  \right>\right) \equiv \bar{D}^S_{\alpha \nu} \delta(t-t')
\end{eqnarray}
\noindent Note $\bar D^S_{\alpha \nu}$ is always real and symmetric. Here, we assume that the electronic
Hamiltonian $\hat H$ is real-valued. 

Eqs. \ref{eq:friction2}-\ref{eq:FFc} are completely general: they express the friction and random force that correspond to the BO approximation, in or out of equilibrium. 
Note that Eqs. \ref{eq:friction2}-\ref{eq:corrTimeQCLE} require taking a full electronic trace, whereas Eq. \ref{eq:mf}  is expressed in terms of single electronic orbitals.  In the SM, we prove that Eq. \ref{eq:friction2} in fact reduces to Tully's expression (Eq. \ref{eq:mf}) at equilibrium if there are no electron-electron interactions.
At this point, there is no simple relationship between $\gamma_{\alpha \nu}$ and $\bar D^S_{\alpha \nu}$.


\textit{Equilibrium, the second fluctuation-dissipation theorem and the symmetry of the friction.---} 
At equilibrium, the steady state electronic density matrix is $\hat \rho_{ss}=e^{-\hat{H}/k_B T}/Z$, where $Z$ is the corresponding normalization factor or  partition function:  $Z=tr_e (e^{-\hat H/k_B T})$.  In the SM (Sec. II), we show that: 
\begin{eqnarray} \label{eq:suhl}
\gamma_{\alpha \nu}   
&=& \frac{1}{ k_B T}
\bar D^S_{\alpha \nu}   
\end{eqnarray}

Thus, the second fluctuation-dissipation theorem is satisfied at equiliibrium. Since $\bar{D}^S_{\alpha \nu}$ is symmetric, $\gamma_{\alpha \nu}$ is also symmetric along $\alpha$ and $\nu$ at equilibrium. 

\textit{Anderson-Holstein model.---} To establish the importance of electron-electron interactions,
we will now calculate the electronic friction for the Anderson-Holstein (AH) model, 
\begin{eqnarray}
\hat H_{AH} &=& \hat H_A + \hat H_{osc} \\
\hat H_A & = & E(x) \sum_{\sigma} \hat d^+_\sigma \hat d_\sigma  +  U \hat n_\uparrow \hat n_\downarrow  +   \sum_{k\sigma} \epsilon_k \hat c_{k\sigma}^+ \hat c_{k\sigma}   \nonumber \\ 
& & +\sum_{k\sigma} V_k  (\hat d^+_\sigma \hat c_{k\sigma} + \hat c_{k\sigma}^+ \hat d_\sigma) \\  
\hat H_{osc} & = & \frac12  \hbar\omega (x^2+p^2) 
\end{eqnarray}
Physically, the AH model represents an electronic impurity $d$ sitting near a metal surface and coupled to a vibrating oscillator ($x$). The impurity can filled with an electron of up or down spin, and so 
$\sigma=\uparrow, \downarrow$ indicates spin. 
The oscillator is a vibrational degree of freedom and feels a different force depending on the occupation of the impurity, $E(x) \equiv E_d + \sqrt{2} g x$.  Note we have defined $x$ and $p$ to be in dimensionless units.

To understand how the motion of the oscillator is perturbed by the fluctuating charge of the impurity, we will calculate the electronic friction.  Of course, to apply Eq. \ref{eq:friction2}, we must diagonalize $\hat H_A$, which will be done via a numerical renormalization group (NRG) calculation \cite{nrgreview}.  We take the wide band approximation, such that the hybridization function $\Gamma \equiv 2\pi \sum_k V_k^2 \delta(\epsilon-\epsilon_k) $ is assumed to be independent of energy.  We leave all details of the calculation to the SM (Sec. III and IV), and show results below.

\begin{figure}[htbp] 
   \centering
   \includegraphics[width=3.5in]{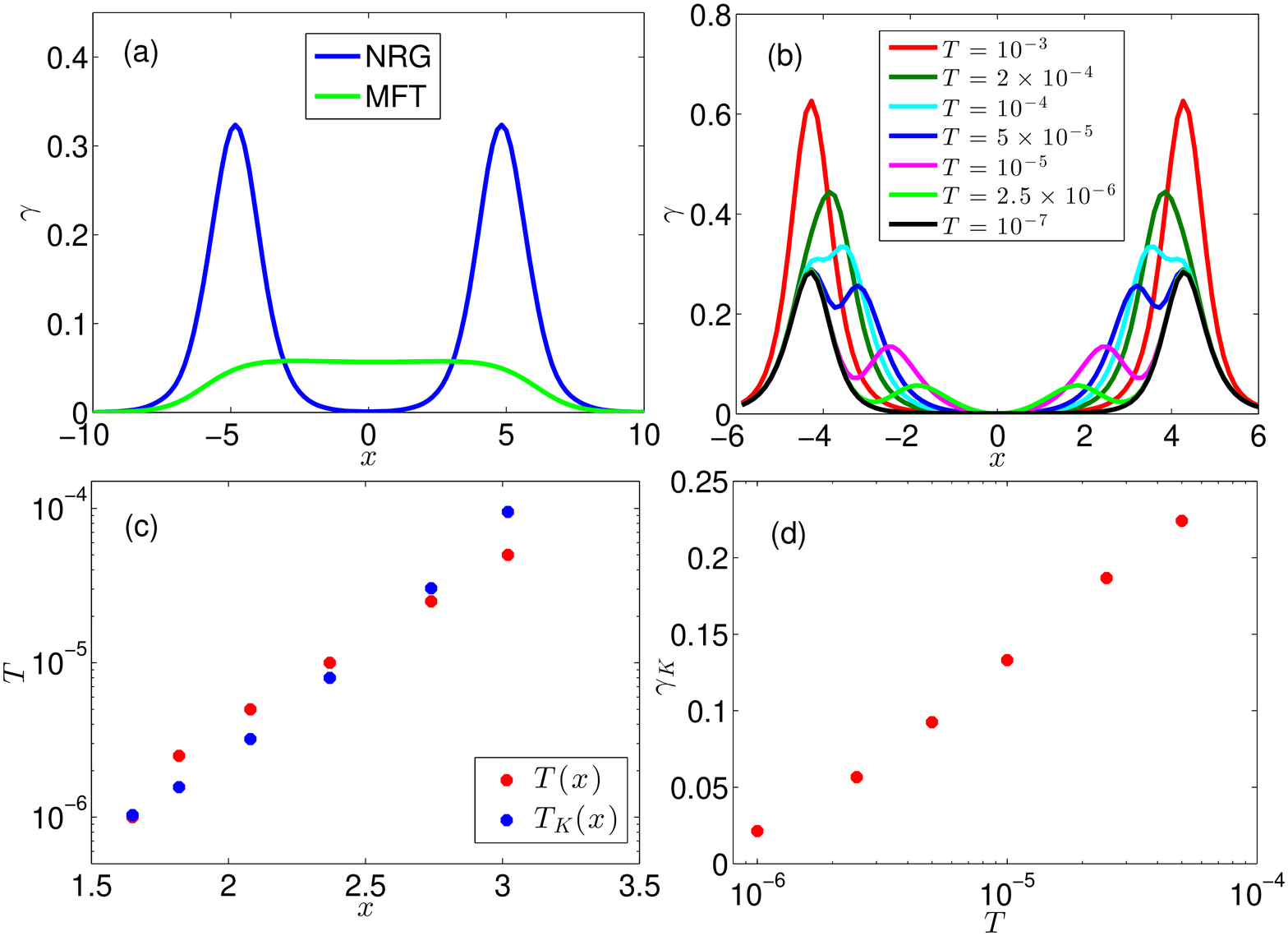} 
   \caption{(a) Electronic friction as function of position $x$ according to both NRG and MFT  \cite{vonOppenJCP2013,PhysRevLett.116.217601} calculations at temperature T=0.005. Note that MFT fails to recover two peaks in the friction.  (b)  Electronic friction according to NRG at low temperature; note that the two peaks in friction become four peaks in friction at low temperature. (c)The Kondo temperature $T_K(x)$ as a function of position and the physical temperature $T(x)$ for which find a peak in friction at position $x$.  Note that these two temperatures are in rough agreement, as predicted by Langreth \cite{LangrethPRB1998}.  (d) The height of the Kondo peak $\gamma_{K}$ as a function of temperature; note that these peaks decrease exponentially and vanish at zero Kelvin, in disagreement with Ref. \cite{LangrethPRB1998}.  Other parameters $U=0.1$, $\Gamma=0.01$, $E_d=-0.05$, $g=0.0075$, bandwidth $D=1$. We have set $k_B=\hbar=1$. }
   \label{fig:1}
\end{figure}

In Fig. \ref{fig:1}(a), we compare the electronic friction as calculated from NRG versus the electronic friction as calculated with mean-field theory (MFT), Eq. \ref{eq:mf}, which is commonly used to treat the Anderson model \cite{vonOppenJCP2013}.  
We study the case of a large repulsion U.  Whereas NRG predicts
two peaks in the electronic friction -- where there is a resonance  of electron attachment/detachment with the Fermi level of the metal $\epsilon_F$ (i.e.  $E_d + \sqrt{2} g x =0$ and $E_d + \sqrt{2}gx+ U = 0 $, where we have set $\epsilon_F=0$)--
MFT predicts only a very broad plateau in friction.
We have attempted to reconstruct these two peaks by manipulating the multiple (broken-symmetry) mean-field solutions of the Anderson-Holstein model, but we have so far been unable to qualitatively match the correct answer.

Beyond the formation of two peaks, even more  interesting feature arises from NRG at still lower temperatures, where Kondo physics now shows a signature. In Fig. \ref{fig:1}(b), we now show that, below $T = 1 \times 10^{-4}$, 
the electronic friction exhibits two additional peaks for a total of {\bf four} peaks. Previously, in Ref. \cite{LangrethPRB1998}, Plihal and Langreth argued
that such new peaks might arise from Kondo resonances.   In other words, we might expect to find peaks in electronic friction at those
positions in space for which the Kondo temperature is equal to the system temperature.  Here, we define the Kondo temperature to be:
 $ T_K(x) = D \exp(- 2\pi |E(x) | |E(x) + U|/U/\Gamma)$, where  $D$ is the bandwidth.
Thus, for Fig. \ref{fig:1}(c), we plot both $T_K(x)$ and the actual $T(x)$, i.e. the temperature at which one finds (with NRG)
a peak in friction at position $x$.  Note that,  
as the data shows, the relevant Kondo temperature $T_K$ is close to the physical temperature $T$, though the agreement is not perfect.

Let us now establish, however,  that Ref. \cite{LangrethPRB1998} (which is based on the non-crossing approximation (NCA) and the case of infinite $U$) 
may not be entirely applicable for our data. 
First, in contrast with Ref. \cite{LangrethPRB1998}, we observe that the
width of the frictional Kondo peak does not 
decrease with the temperature.
Second, again in contrast with Ref. \cite{LangrethPRB1998},  we find that 
 the frictional peaks associated with a 
Kondo resonance disappear as the temperature decreases to zero (rather than increase).
The height  of these peaks is plotted in Fig. \ref{fig:1}(d) and appears
to decrease exponentially with temperature.
Future pencil and paper work will be required to understand these features.


\textit{Conclusions.---} We have derived a universal expression for the electronic friction as experienced by a set of classical nuclear particles coupled to a manifold of quantum mechanical electronic DoF's.  The key equations are Eqs. \ref{eq:friction2}, \ref{eq:frictionGA} and \ref{eq:corrTimeQCLE}.  The derivation is simple and, in the same spirit as a Mori-Zwanzig projection:  Assuming all  dynamics follow the quantum-classical Liouville equation, we simply make an adiabatic approximation and isolate fluctuations around a slow variable.  Our final expression is quite general, insofar as it applies in/out of equilibrium and with an arbitrary electronic Hamiltonian; at equilibrium, our work validates
the Suhl's ``bootstrap'' conjecture \cite{PhysRevB.11.2122}.

Looking forward, Eq. \ref{eq:friction2} is demanding to apply because, without any further approximations,
these equations require the full diagonalization the electronic Hamiltonian in terms
of the many-body electronic states
(much like the Meir-Wingreen formula \cite{PhysRevLett.68.2512}).  
However, using numerical renormalization group (NRG) theory, we have now shown 
how to calculate electronic friction tensors exactly for small model problems. 
Thus,  extending the work of von Oppen \textit{et al} \cite{beilstein},
 one can now study  
model Hamiltonians and learn how nuclear motion near metal surfaces will be effected by electron-electron correlation, either
with or without a current through the molecule.  For example, in the present letter, we have studied the Anderson-Holstein model and shown that the usual electronic friction
expression (with mean-field theory) yields qualitatively wrong features; when we account for electronic friction,
we find multiple peaks, including two associated with Kondo physics.
Beyond model problems, even if we cannot use NRG to diagonalize the Hamiltonian, 
there are currently several research groups that are seeking to calculate approximate energies for extended systems
beyond mean-field theory (MFT).  For instance, within the condensed matter world, there is currently a big push to calculate
correlated electron attachment/detachment energies with GW \cite{PhysRevB.34.5390} and optical excitations with the Bethe-Salper Equation, 
all for periodic systems \cite{deslippe2012berkeleygw, PhysRevB.62.4927}. Furthermore, within the chemistry world, there has been recent work to calculate ground state coupled-cluster (CC) and 
excited state equation of motion CC energies for periodic systems \cite{mcclain2017gaussian, mcclain2016spectral}.  All of these methods can be
used to calculate electronic friction through 
Eq. \ref{eq:frictionGA}  going beyond MFT and thus 
improve our understanding of nuclear motion in the condensed phase.

Finally, one pressing question remains regarding the validity of electronic friction: just like the BO approximation,
the adiabatic approximation that we make  (in the SM) to move from Eq. 38 to Eq. 39 
is uncontrolled and without a unique small parameter.   On the one hand, Tully has argued
that electronic friction fails for NO scattering off of gold \cite{shenvi:2009:science}. On the other hand, 
while one would certainly expect electronic friction to fail
in the nonadiabatic Marcus electron transfer regime, in fact
we recently showed \cite{ouyang2016dynamics} that, with just a little bit of external nuclear friction, 
electronic friction recovers Marcus theory.  From Ref. \cite{ouyang2016dynamics}, 
the only lesson we have so far learned is that 
Langevin dynamics with electronic friction will fail if truly excited state dynamics appear.
In this case, one possible path forward is to include memory effects with a non-Markovian Fokker-Planck equation \cite{smith1993electronic} (see the SM); another approach is to employ surface hopping techniques
that can include frictional effects \cite{paperII, mulFrictionPaperJCP2016, doi:10.1021/acs.jctc.7b00094}.
In the future, it will be essential to further investigate when and how such excited state dynamics occurs,
and given how important are electronic-electronic interactions for identifying curve crossings,
the present paper takes an important step forward by unambiguously identifying
the correct, universal electronic friction tensor accompanying all Born-Oppenheimer dynamics.

This material is based upon work supported by the (U.S.) Air Force Office of Scientific Research (USAFOSR) PECASE award under AFOSR Grant No. FA9950-13-1-0157.
J.E.S. gratefully acknowledges support from the
Stanford PULSE Institute and a John Simon Guggenheim
Memorial fellowship.

\end{document}